\documentclass[journal=jctcce,manuscript=article,layout=traditional]{achemso} 
\usepackage[T1]{fontenc}
\usepackage[version=3]{mhchem}
\usepackage[dvipsnames]{xcolor}
\usepackage{graphicx,subcaption}
\usepackage{amssymb}
\usepackage{overpic}
\usepackage{braket}
\usepackage{soul}
\usepackage{bm}
\usepackage{hyperref}
\usepackage{enumerate}
\hypersetup{pdfborderstyle={/S/U/W 0.5}}

\newcommand{\pp}[2]{\frac{\partial{#1}}{\partial{#2}}}
\newcommand{\grad}[1]{\nabla_{\bm{#1}}}
\author{Tian Qiu}
\email{tianq@princeton.edu}
\affiliation{Department of Chemistry, Princeton University, Princeton, NJ 08540, USA.}
\author{Joseph E. Subotnik}

\affiliation{Department of Chemistry, Princeton University, Princeton, NJ 08540, USA.}
\title[]{Fast Methods For Multisite Charge Transfer Processes I: Constrained, State Averaged CASSCF(1,M) and CASSCF(2M-1,M) Simulations }

\begin{document}
\maketitle

\begin{abstract}
We design a 
dynamically-weighted state-averaged constrained CASSCF to treat \ul{e}lectrons or \ul{h}oles moving between $n$  molecular fragments (where $n$ can be larger than 2). Within such a so-called eDSCn/hDSCn approach, we consider configurations that are mutually single excitations of each other, and we apply a generalized set of constraints to tailor the method for studying charge transfer problems. The constrained optimization problem is efficiently solved using a DIIS-SQP algorithm, thus maintaining computational efficiency. We demonstrate the method for a  finite Su-Schrieffer-Heeger (SSH) chain, successfully reproducing the expected exponential decay of diabatic couplings with distance. When combined with a gradient, the current  extension immediately enables efficient nonadiabatic dynamics simulations of complex multi-state charge transfer processes.
\end{abstract}

\section{Introduction}
Electron transfer through multiple charge centers play a fundamental role in complex physical and chemical processes. For example, multi-site electron transfer governs the rates of electrocatalytic reactions \cite{Koper2013,Jiao:2015:electrocatalysts}, determines the efficiency of multi-junction solar cells \cite{Yamaguchi2021}, and controls  charge transport in molecular electronics \cite{nitzan:2001:review,ALBINSSON2008,Newton:2003:electron_transfer}.
Obviously, understanding the coupling  between electronic states is crucial, as it is known that $(i)$ in the limit of weak coupling, there is incoherent sequential charge transfer between sites and (ii)  in the limit of strong coupling, one finds very fast band transport.\cite{Tretiak:2002:electron_excitation_conjugate,Newton:2003:electron_transfer,Matsika:2011:conical_intersection,Su:2018:fractional_spin_dft,Casanova:2021:strong_correlation}  These rules of thumb hold in general, and can be ascertained easily for models of noninteracting electronic Hamiltonians. 
That being said, however, if we seek to go beyond model Hamiltonians and investigate multisite electron transfer (and ideally electron transport) within an {\em ab initio }context,  we face several obstacles.
First, the presence of multiple  states introduces competing pathways for charge transfer (CT) and increased quantum effects, such that  one cannot treat CT between each pair of charge centers separately.
Second, to account for electron-phonon coupling, a meaningful theoretical description must switch smoothly between charge-localized state and delocalized states across multiple sites that may change dramatically with molecular geometry, external fields, and chemical environment. 
Third,  it is critical that we recover balanced electron correlation energies for electrons in different environments while also maintaining efficiency and computational accessibility for dynamics simulations; in particular,
it is essential that one treat multiple electronic configurations on an equal footing.


To that end, single-reference methods such as Hartree-Fock (HF) or Density Functional Theory (DFT) are clearly not the appropriate methods of choice, as they fail to capture the multi-reference nature of CT processes; Time-Dependent DFT (TD-DFT) has emerged as a widely-adopted approach for modeling charge transfer processes,\cite{CarterFenk:2021:CT_TDDFT,Dai:2022:excitation} but the latter treats excited states and the ground state in an unbalanced fashion  (and cannot recover the correct topology of a crossing\cite{martinez:2006:ci_topology_wrong}).  Recent work in our group\cite{Athavale2021} (and  analogous work from other groups\cite{Li:2014:ci_tda,Xu:2025:cis_tda})  to restore balance can indeed work for electron transfer between two sites but not has not yet been generalized to multiple site problems.  Multi-reference methods such as Complete Active Space Self-Consistent Field (CASSCF), Multi-Reference Configuration Interaction (MRCI),\cite{Szalay:2011:mrci,Lischka:2018:MR} and density matrix renormalization group (DMRG) theory are natural candidates for future work as they can provide accurate descriptions -- but these methods do suffer from rapid computational scaling with the size of the system and the active space. 
Moreover, it is worth noting that working with CASSCF becomes more and more difficult for systems with more and and more electronic states not only as far as memory costs, but also because, in such contexts, one often finds multiple solutions and discontinuous potential energy surfaces.

With this background in mind, in a recent article
\cite{Qiu:2024:dsc,Qiu:2024:diis}, we took an initial step toward solving the problem above by introducing an electron/hole dynamically-weighted state-averaged constrained CASSCF (eDSC/hDSC) method for studying two-state CT processes in radical systems. By constraining the active orbital space to project equally onto donor and acceptor fragments, the eDSC/hDSC approach was able to generate nearly globally smooth, accurate potential energy surfaces at approximately twice the cost of restricted open-shell Hartree-Fock (ROHF) calculations. In addition, the method was able to correct  describe not only weakly but also strongly coupled CT processes (where methods like constrained DFT (cDFT) fail).  That being said,  the method delineated in Ref. \citenum{Qiu:2024:dsc} was restricted to a two-site system; and given the discussion above, there would be an enormous benefit (with many applications) if one could extend the eDSC/hDSC approach so as to treat multi-state electron or hole transfer problems in an efficient manner.

To that end, in the present paper, we will show that one can indeed extend the eDSC/hDSC methodology to treat CT through $n$ multiple charge centers / molecular fragments, resulting in an algorithm that we will refer to as eDSCn/hDSCn. 
Just as for eDSC/hDSC\cite{Qiu:2024:diis}, the algorithm is able to efficiently
treat multiple electronic states and constraints efficiently by using a DIIS-SQP solver for the constrained optimization problem. At the end fo the day, we believe the method opens the door to efficient nonadiabatic dynamics simulations of complex multi-state CT processes, which has broad applications in study of energy conversion, molecular electronics, and biological electron transfer.

An outline of the paper is as follows. In Sec. \ref{sec:review}, we briefly review the eDSC/hDSC method. In Sec. \ref{sec:extension}, we extend the formalism to the case of more than two molecular fragments and configurations. In Sec. \ref{sec:diis_sqp}, we derive the working equations for solving the extended formalism with the DIIS-SQP algorithm. In Sec. \ref{sec:general_constraint}, we introduce a generalized constraint formulation that provides flexibility in treating different problems. In Sec. \ref{sec:result}, we present the results for the finite Su–Schrieffer–Heeger chains. In Sec. \ref{sec:discussion}, we discuss the physical meaning of a convergence trick applied in the algorithm. In Sec. \ref{sec:conclusion}, we conclude this manuscript and provide outlook for future research directions.

\section{Review of eDSC/hDSC}\label{sec:review}
Before describing the eDSCn/hDSCn algorithm detail, let us review the two-state eDSC/hDSC approach from  Ref. \citenum{Qiu:2024:dsc}.  For the two state problem, we define the target function for eDSC (exciting the electron)/hDSC (exciting the hole) as
\begin{align}
    E_{\rm tot} &= w_1E_1 + w_2E_2,\label{eq:e_tot}
\end{align}
where $E_1$ and $E_2$ are energies of two restricted-open-shell electron configurations that both have one unpaired electron. For eDSC, the two configurations are
\begin{align}
    \ket{\Psi_1} &= \ket{1\bar{1},2\bar{2},...N\bar{N},N+1},\label{eq:config_e_1}\\
    \ket{\Psi_2} &= \ket{1\bar{1},2\bar{2},...N\bar{N},N+2},\label{eq:config_e_2}
\end{align}
and for hDSC, the two configurations are
\begin{align}
    \ket{\Psi_1} &= \ket{1\bar{1},2\bar{2},...,N\bar{N},N+1},\\
    \ket{\Psi_2} &= \ket{1\bar{1},2\bar{2},...,N,N+1\overline{N+1}}.
\end{align}
$w_1$ and $w_2$ in Eq. \ref{eq:e_tot} are dynamical weights that depend on a ``temperature'' parameter $T$ and the energy difference $\Delta E = E_2 - E_1$, where we assume $E_2 > E_1$. For the two-state problem with only one active electron (and effectively zero correlation energy) it is natural to define the weights as:
\begin{align}
    w_1(\Delta E) &= 1 - \frac{1-e^{-\Delta E/T}}{2\Delta E/T},\\
    w_2(\Delta E) &= \frac{1-e^{-\Delta E/T}}{2\Delta E/T}.
\end{align}
The differential of Eq. \ref{eq:e_tot} is
\begin{align}
    dE_{\rm tot} &= w_1'dE_1 + w_2'dE_2
\end{align}
where
\begin{align}
    w_1' &= (1-\frac{1}{2}e^{-\Delta E/T}),\label{eq:w_1}\\
    w_2' &= \frac{1}{2}e^{-\Delta E/T}.\label{eq:w_2}
\end{align}
Note that $w_1'\neq \pp{w_1}{\Delta E}$.

At the heart of the eDSC/hDSC is the need to suppress any and all local excitations; in so doing, one can find nearly globally smooth surfaces as relevant for charge transfer.  In order to achieve such a suppression, we impose a constraint that the vector space spanned by active orbitals (i.e., orbitals $N+1$ and $N+2$ for eDSC and orbitals $N$ and $N+1$ for hDSC) projects equally onto the vector space spanned by orbitals from the donor and acceptor molecular fragments, i.e.,
\begin{align}
    {\rm Tr}(\hat{P}^D\hat{P}^{\rm active} - \hat{P}^A\hat{P}^{\rm active}) = 0\label{eq:q_formal}.
\end{align}
where $\hat{P}^D$ and $\hat{P}^A$ are vector space spanned by orbitals from the donor and acceptor molecular fragments, respectively, and $\hat{P}^{\rm active}$ is the vector space spanned by active orbitals. Eq. \ref{eq:q_formal} defines a constrained optimization problem and one can write the Lagrangian formally as
\begin{align}
    \mathcal{L} = w_1E_1 + w_2E_2 - \lambda {\rm Tr}(\hat{P}^D\hat{P}^{\rm active} - \hat{P}^A\hat{P}^{\rm active}) \label{eq:lag_formal}
\end{align}
where $\lambda$ is the Lagrange multiplier. To solve this problem efficiently, in  Ref. \citenum{Qiu:2024:dsc}, we applied the sequential quadratic programming (SQP) method and later introduced a DIIS-SQP algorithm in Ref. \citenum{Qiu:2024:diis} to further and dramatically reduce the computational cost to only twice the cost of Restricted Open-shell Hartree Fock (ROHF) calculation. 

In the present paper, our goal will now be to extend this theory to study systems with more than two charge centers and active orbitals. In a companion paper, we will further derive the relevant analytic nuclear gradient of adiabatic energies and the derivative coupling.

Before proceeding, we remind the reader that the primary purpose of eDSCn/hDSCn is to study electron/hole transfer in a radical system and we  consider only electronic configurations that are mutually single excitations of each other.  In other words, we seek the absolutely most efficient and streamlined active space for coupled nuclear-electronic charge transfer simulations, and generating such a platform does necessarily leave a risk that we will  ignore important electron-electron correlations. That being said, for large scale simluations, strong approximations are imperative;  all of the  configurations for eDSCn/hDSCn are defined explicitly in Sec. \ref{sec:extension}.

\section[fragments and configurations]{Extension to $M^f$ Fragments with $M^c$ Configurations}\label{sec:extension}
To probe $M^c$ configurations in an odd-number electron system, as above, it is natural define the total energy as the dynamical weighted sum of energies from all configurations, namely,
\begin{align}
    E_{\rm tot} = \sum_{j=1}^{M^c} w_jE_j,\label{eq:E_tot}
\end{align}
where $E_j$ is the energy of the $j$th electronic configuration, $\ket{\Psi_j}$, and $w_j$ is the corresponding weighting factor. Similar to eDSC/hDSC, we will also make a distinction between electron transfer and hole transfer and refer electron transfer as eDSCn and hole transfer as hDSCn. Importantly, the electron configurations considered depend on whether we study the electron transfer or hole transfer. Specifically, for eDSCn, let the system have $2N+1$ electrons, the $j$th electronic configuration is ($j=1\cdots M^c$)
\begin{align}
    \ket{\Psi_j} = \ket{1\bar{1},2\bar{2},...,N\bar{N},N+j},\label{eq:edscn_wfc}
\end{align}
which consists the complete active space for one electron in $M^c$ orbitals, often referred to as CAS$(1,M^c)$ in literature. For hDSCn, let the system have $2N-1$ electrons, the electronic configurations we considered are (again, $j=1\cdots M^c$)
\begin{align}
    \ket{\Psi_j} = \ket{1\bar{1},2\bar{2},...N-j,\overline{N-j},N-j+1,N-j+2,\overline{N-j+2},...,N\bar{N}},\label{eq:hdscn_wfc}
\end{align}
which consists the complete active space for $2M^c-1$ electron in $M^c$ orbitals, or CAS$(2M^c-1,M^c)$. For both eDSCn and hDSCn, we will refer to the singly occupied orbital in each configuration as the active orbital.

The next step is to develop the appropriate constraints for systems with more than 2 molecular fragments. To suppress local excitations in a system where the charge can travel through $M^f$ centers, we find that the best approach is to study $M^c = M^f$ configurations and apply the constraints that the vector space spanned by all active orbitals projects equally onto each molecular fragments, i.e.,
\begin{align}
    {\rm Tr}\left(\hat{P}^1\hat{P}^{\rm active}\right) = {\rm Tr}\left(\hat{P}^2\hat{P}^{\rm active}\right) = \cdots = {\rm Tr}\left(\hat{P}^{M^f}\hat{P}^{\rm active}\right).\label{eq:constraints}
\end{align}
where $\hat{P}^{j}$ is the vector space spanned by atomic orbitals from the $j$th molecular fragment and $\hat{P}^{\rm active}$ is the vector space spanned by all active orbitals. Thus, for a system with $M^f$ fragments (i.e., charge centers), we have $M^f-1$ constraints:
\begin{align}
    {\rm Tr}\left((\hat{P}^j-\hat{P}^{j+1})\hat{P}^{\rm active}\right) = 0,\ j=1\cdots M^f-1.\label{eq:constraints_v2}
\end{align}

Finally, the weighting factors $w_j$ in Eq. \ref{eq:E_tot} need to be defined in a slightly different way than Eqs. \ref{eq:w_1}-\ref{eq:w_2}. We propose the following form (assuming $\ket{\Psi_1}$ has the lowest energy)
\begin{align}
    w_j = \frac{a_j}{\sum_j a_j},
\end{align}
where
\begin{align}
    a_j &= \frac{1-e^{-\Delta E_j/T}}{\Delta E_j/T},\\
    \Delta E_j &= E_j - E_1.
\end{align}
Note that $a_1 = \lim_{\Delta E\rightarrow 0}\frac{1-e^{-\Delta E/T}}{\Delta E/T} = 1$. With this definition, the first differential of $E_{\rm tot}$ can be written as
\begin{align}
    dE_{\rm tot} = \sum_j w_j'dE_j,
\end{align}
where
\begin{align}
    w_{j>1}' &= \frac{1}{\sum_j a_j}\left(e^{-\Delta E_j/T}-\frac{E_{\rm tot}-E_1}{\Delta E_j}\left(e^{-\Delta E_j/T}-a_j\right)\right),\label{eq:wp_2}\\
    w_1' &= 1-\sum_{j>1}w_j'.\label{eq:wp_1}
\end{align}
At this point, one can write a formal Lagrangian for the constrained optimization problem
\begin{align}
    \mathcal{L} = \sum_{j=1}^{M^c} w_jE_j - \sum_{j=1}^{{M^f}-1}\lambda_j{\rm Tr}\left((\hat{P}^j-\hat{P}^{j+1})\hat{P}^{\rm active}\right).\label{eq:lagrangian_formal}
\end{align}

\section{Solving Constrained Optimization Problem with DIIS-SQP Framework}\label{sec:diis_sqp}
At the heart of a eDSCn/hDSCn calculation is a constrained optimization problem represented by the Lagrangian in Eq. \ref{eq:lagrangian_formal}. The framework applied here is an extension to the approach in Ref. \citenum{Qiu:2024:diis}. For maximum clarity, we will  deal with eDSCn in Sec. \ref{sec:edsc}; we will review  necessary revisions for solving hDSCn problems in Sec. \ref{sec:hdsc}.

\subsection{Case of eDSCn}\label{sec:edsc}
Let $\bm{P}^0$ be the density matrix of orbitals that are always doubly occupied, i.e. the one-electron molecular orbitals $\phi_1,\phi_2,\cdots,\phi_{N}$.  Let $\bm{P}^{j>0}$ be the density matrix of singly occupied orbitals in configuration $\ket{\Psi_j}$, i.e., $\phi_{N+j}$. The energy of the $j$th configuration can be expressed as
\begin{align}
    E_j = \frac{1}{2}{\rm Tr}\left[(\bm{H}^0+\bm{F}^{j\alpha})\bm{P}^0\right] + \frac{1}{2}{\rm Tr}\left[(\bm{H}^0+\bm{F}^{j\beta})(\bm{P}^0+\bm{P}^j)\right]\label{eq:eje}
\end{align}
where $\bm{H}^0$ is the one-electron Hamiltonian, $\bm{F}^{j\alpha}$, $\bm{F}^{j\beta}$ are spin up and down Fock matrices built from spin densities matrices that $\bm{P}^{j\alpha} \equiv \bm{P}^0$, $\bm{P}^{j\beta} \equiv \bm{P}^0+\bm{P}^j$, respectively. Next, let $\bm{Q}^j$ represents the projection operator difference between fragment $j$ and $j+1$, i.e.,
\begin{align}
    \bm{Q}^j = \hat{P}^j-\hat{P}^{j+1}\label{eq:qj},
\end{align}
such that our constraints can be represented most simply as
\begin{align}
    {\rm Tr}\left[\bm{Q}^j\bm{P}^{\rm active}\right] = 0,\label{eq:trQP}
\end{align}
where
\begin{align}
    \bm{P}^{\rm active} = \sum_{j=1}^{M^c}\bm{P}^j.
\end{align}
One can rewrite the Lagrangian in Eq. \ref{eq:lagrangian_formal} using the density matrices $\bm{P}^j$ (here we assume an orthonormal basis)
\begin{align}
    \mathcal{L} = &\sum_{j=1}^{M^c} w_jE_j - \sum_{j=0}^{M^c} {\rm Tr}\left[\bm{\epsilon}^j((\bm{P}^j)^2-\bm{P}^j)\right]-\sum_{j>i\geq 0}^{M^c}{\rm Tr}\left[\bm{\sigma}^{ij}(\bm{P}^i\bm{P}^j+\bm{P}^j\bm{P}^i)\right]\nonumber\\
    -&{\rm Tr}\left[\left(\sum_{j=1}^{M^f-1}\lambda_j\bm{Q}^j\right)\left(\sum_{j=1}^{M^c}\bm{P}^j\right)\right],\label{eq:lagrangian}
\end{align}
where $\bm{\epsilon}^j$, $\bm{\sigma}^{ij}$, and $\lambda_j$ are Lagrangian multipliers. Note that $\bm{\epsilon}^j$ and $\bm{\sigma}^{ij}$ are symmetric matrices and with the same dimension as $\bm{P}^j$. In Eq. \ref{eq:lagrangian}, $\lambda_j$ enforces the charge transfer constraints we introduced in Eq. \ref{eq:constraints}, $\bm{\epsilon}^j$ ensures that the density matrices $\bm{P}^j$ are idempotent, and $\bm{\sigma}^{ij}$ ensures that $\bm{P}^i$ and $\bm{P}^j$ are orthogonal for $i\neq j$ (see Ref. \citenum{Qiu:2024:diis}).
We will now outline how to solve the 
constrained optimization problem in Eq. \ref{eq:lagrangian} in a series of seven steps.

\begin{enumerate} 
\item To begin with, one can write
\begin{align}
    \left(\grad{P^0}\mathcal{L}\right)\bm{P}^0 &= \sum_{j=1}^{M^c} w_j'(\bm{F}^{j\alpha}+\bm{F}^{j\beta})\bm{P}^0-\bm{P}^0\bm{\epsilon}^0\bm{P}^0 - \sum_{j=1}^{M^c}\bm{P}^j\bm{\sigma}^{0j}\bm{P}^0 = \bm{0},\label{eq:dldp0}\\
    \left(\grad{P^k}\mathcal{L}\right)\bm{P}^k &= \left(w_k'\bm{F}^{k\beta}-\sum_{j=1}^{M^f-1}\lambda_j\bm{Q}^j\right)\bm{P}^k - \bm{P}^k\bm{\epsilon}^k\bm{P}^k - \sum_{\substack{j=0\\j\neq k}}^{M^c} \bm{P}^j\bm{\sigma}^{kj}\bm{P}^k = \bm{0}, k\neq 0.\label{eq:dldpk}
\end{align}
Note that in the equations above, we define $\bm{\sigma}^{k>j} = \bm{\sigma}^{jk}$ to keep the equations formally symmetric. ($\bm{\sigma}^{kj}$ is defined only for $k<j$ in Eq. \ref{eq:lagrangian})

\item Let us define
\begin{align}
    \bm{M}^0 &= \sum_{j=1}^{M^c} w_j'(\bm{F}^{j\alpha}+\bm{F}^{j\beta}),\label{eq:m0edscn}\\
    \bm{M}^{j>0} &= w_j'\bm{F}^{j\beta}\label{eq:mjedscn},\\
    \bm{Q}^0 &= \sum_{j=1}^{M^f-1}\lambda_j\bm{Q}^j.
\end{align}
At this point, if we add Eqs. \ref{eq:dldp0} and \ref{eq:dldpk} for all $k$ together, we find: 
\begin{align}
    \bm{M}^0\bm{P}^0 +\sum_{j=1}^{M^c}\left(\bm{M}^j-\bm{Q}^0\right)\bm{P}^j = \sum_{j=0}^{M^c}\bm{P}^j\bm{\epsilon}^j\bm{P}^j + \sum_{\substack{j=0\\k=0\\j\neq k}}^{M^c}\bm{P}^j\bm{\sigma}^{kj}\bm{P}^k.\label{eq:edscn_together}
\end{align}

\item
Since $\bm{P}^j$, $\bm{\epsilon}^{j}$, and $\bm{\sigma}^{kj}$ are all symmetric matrices, the R.H.S. of Eq. \ref{eq:edscn_together} is also symmetric, which suggests that the L.H.S. of Eq. \ref{eq:edscn_together} must be symmetric at the solution of the constrained optimization problem. In other words, one can use this property to measure the deviation or error from the solution and define a DIIS error vector accordingly:
\begin{align}
    \bm{V}_{\rm DIIS} = [\bm{M}^0,\bm{P}^0] + \sum_{j=1}^{M^c} [\bm{M}^j-\bm{Q}^0,\bm{P}^j]\label{eq:diis_err_vec}
\end{align}
Here,  $[\bm{A},\bm{B}] \equiv \bm{AB}-\bm{BA}$ is the commutator.

\item Following Ref. \citenum{Qiu:2024:diis}, one can establish a DIIS procedure for the optimization of Eq. \ref{eq:lagrangian} provided that Eqs. \ref{eq:dldp0}-\ref{eq:dldpk} can be solved for fixed quantities of $w_j',\bm{F}^{j\alpha},\bm{F}^{j\beta}$, or equivalently, $\bm{M}^j$. In particular, it  can be verified that solving Eqs. \ref{eq:dldp0}-\ref{eq:dldpk} with fixed $\bm{M}^j$ is equivalent to minimizing the auxiliary function 
\begin{align}
    E_{\rm aux}(\{\bm{P}^j\}) = \sum_{j=0}^{M^c}{\rm Tr}\left[\bm{M}^j\bm{P}^j\right]
\end{align}
under the same set of constraints in Eq. \ref{eq:lagrangian}. 

\item To work with the smallest parameter space possible, one can absorb the orthonormality constraints into the parametrization by writing 
\begin{align}
    \bm{P}^j &= \bm{C}\bm{N}^j\bm{C}^\top,\label{eq:pj_from_C}
\end{align}
where $\bm{C}$ represents molecular orbital coefficients (in an orthonormal basis, $\bm{C}$ is a unitary matrix) and $\bm{N}^j$ is an occupation matrix, and parameterize $\bm{C}$ locally by an anti-symmetric matrix $\bm{A}$ (rotation generator), i.e.,
\begin{align}
    \bm{C} = \bm{C}_{0}e^{\bm{A}}.
\end{align}
With this transformation, one can build an auxiliary Lagrangian
\begin{align}
    \mathcal{L}_{\rm aux} = \sum_{j=0}^{M^c}{\rm Tr}\left[e^{-\bm{A}}\bm{C}_0^\top\bm{M}^j\bm{C}_0e^{\bm{A}}\bm{N}^j\right] - \sum_{k=1}^{M^f-1}\lambda_k{\rm Tr}\left[e^{-\bm{A}}\bm{C}_0^\top\bm{Q}^k\bm{C}_0e^{\bm{A}}\sum_{j=1}^{M^c}\bm{N}^j\right],\label{eq:l_aux}
\end{align}
where $\lambda_k$ are Lagrangian multipliers.

\item The constrained minimization problem in Eq. \ref{eq:l_aux} can be solved with SQP, which requires the gradient of the target function and the constraints w.r.t. $\bm{A}$. The energy term gives (note that $\bm{A}$ is anti-symmetric)
\begin{align}
    \left.\pp{E_{\rm aux}}{A_{pq}}\right|_{\bm{A}=\bm{0}} = 2\left(\sum_{j=0}^{M^c}[\tilde{\bm{M}}^j,\bm{N}^j]\right)_{pq}\ \ ,q>p\label{eq:dEdA}
\end{align}
where 
\begin{align}
    \tilde{\bm{M}}^j=\bm{C}_0^\top \bm{M}^j \bm{C}_0,\label{eq:mtilde}
\end{align}
and the constraint term gives
\begin{align}
    \left.\pp{}{A_{pq}}{\rm Tr}\left[e^{-\bm{A}}\bm{C}_0^\top\bm{Q}^k\bm{C}_0e^{\bm{A}}\sum_{j=1}^{M^c}\bm{N}^j\right]\right|_{\bm{A}=\bm{0}} = 2\left([\tilde{\bm{Q}}^k,\sum_{j=1}^{M^c}\bm{N}^j]\right)_{pq}\ \ ,q>p\label{eq:dQdA}
\end{align}
where $\tilde{\bm{Q}}^k = \bm{C}_0^\top \bm{Q}^k \bm{C}_0$.
Since Eqs \ref{eq:dEdA} and \ref{eq:dQdA} are only valid at $\bm{A}=\bm{0}$, $\bm{C}_0$ needs to be updated in each iteration.

\item For practical optimization problems, it is often desirable to use the diagonal of the Hessian matrix as a preconditioner. One can verify that
\begin{align}
    W_{pq}\equiv\left.\pp{^2E_{\rm aux}}{A_{pq}^2}\right|_{\bm{A}=\bm{0}} &= 2\sum_{j=0}^{M^c}\left(\left(\bm{N}^{j}\right)_{pp}-\left(\bm{N}^{j}\right)_{qq}\right)\left(\left(\tilde{\bm{M}}^{j}\right)_{pp}-\left(\tilde{\bm{M}}^{j}\right)_{qq}\right).
\end{align}
Similar to the algorithm in Ref. \citenum{Qiu:2024:diis}, let $df_{pq}$ and $dg_{pq}^k$ be the gradient of the target function and the $k$th constraint that are required in SQP algorithm, the preconditioner suggests that instead of using Eq. \ref{eq:dEdA} and \ref{eq:dQdA} as the expression for $df_{pq}$ and $dg_{pq}^k$, one can instead use
\begin{align}
    df_{pq} &= 2\left.\left(\sum_{j=0}^{M^c}[\tilde{\bm{M}}^j,\bm{N}^j]\right)_{pq} \right/ |W_{pq}|\label{eq:sqp_df}\\
    dg_{pq}^k &= 2\left.\left([\tilde{\bm{Q}}^k,\sum_{j=1}^{M^c}\bm{N}^j]\right)_{pq} \right/ |W_{pq}|.\label{eq:sqp_dg}
\end{align}
\end{enumerate}

The analysis above highlights the key conceptual points needed for optimization; the remaining procedure is identical to the algorithm proposed in Ref. \citenum{Qiu:2024:diis}. For the convenience of the reader, we provide a flowchart in Sec. \ref{sec:flowchart}.

\subsection{Case of hDSCn}\label{sec:hdsc}
The same procedure as above can be applied to hDSCn after redefining some quantities. To be specific, different from Eq. \ref{eq:eje}, the energy expressions for configurations in the case of hDSCn can be written as
\begin{align}
    E_j = \frac{1}{2}{\rm Tr}\left[(\bm{H}^0+\bm{F}^{j\alpha})\bm{P}^0\right] + \frac{1}{2}{\rm Tr}\left[(\bm{H}^0+\bm{F}^{j\beta})(\bm{P}^0-\bm{P}^j)\right],\label{eq:ejh}
\end{align}
where $\bm{P}^0$ still represents the density matrix from molecular orbitals $\phi_1,\phi_2,\cdots,\phi_{N}$. Note that, though, we define $N$ by counting $2N-1$ total electrons in hDSCn (see Eq. \ref{eq:hdscn_wfc}) rather than $2N+1$ total electrons in eDSCn (see Eq. \ref{eq:edscn_wfc}), such that $\bm{P}^0$ differs when applying eDSCn or hDSCn even to the same molecule with the same number of electrons. $\bm{P}^{j>0}$ still represents the density matrix from singly occupied orbitals in configuration $\ket{\Psi_j}$, but this molecular orbital is now $\phi_{N+1-j}$. $\bm{H}^0$ is still the one-electron Hamiltonian, $\bm{F}^{j\alpha}$, $\bm{F}^{j\beta}$ are spin up and down Fock matrices built from the spin densities  $\bm{P}^{j\alpha} \equiv \bm{P}^0$, $\bm{P}^{j\beta} \equiv \bm{P}^0-\bm{P}^j$. As a result, Eqs. \ref{eq:m0edscn}-\ref{eq:mjedscn} should be replaced by
\begin{align}
    \bm{M}^0 &= \sum_{j=1}^{M^c} w_j'(\bm{F}^{j\alpha}+\bm{F}^{j\beta}),\label{eq:m0hdscn}\\
    \bm{M}^{j>0} &= -w_j'\bm{F}^{j\beta}.\label{eq:mjhdscn}
\end{align}
With these revised definition of $\bm{M}^j$, the rest of the procedure remains the same as eDSCn in Sec. \ref{sec:edsc}.

\subsection{Avoiding the saddle point}\label{sec:saddle}
As discussed in Ref. \citenum{Qiu:2024:diis}, there is one extra symmetry that needs to be addressed to avoid converging to the saddle points during the optimization procedure. Specifically, it can be shown that for eDSCn, the following identity holds:
\begin{align}
    \left(\tilde{\bm{F}}^{j\beta}\right)_{N+j,N+k} = \left(\tilde{\bm{F}}^{k\beta}\right)_{N+j,N+k},\label{eq:f_equal_e}
\end{align}
where
\begin{align}
    \tilde{\bm{F}}^{j\beta} &= \bm{C}^\top \bm{F}^{j\beta}\bm{C}.
\end{align}
Similarly, for hDSCn
\begin{align}
    \left(\tilde{\bm{F}}^{j\beta}\right)_{N+1-j,N+1-k} = \left(\tilde{\bm{F}}^{k\beta}\right)_{N+1-j,N+1-k}.\label{eq:f_equal_h}
\end{align}
To avoid the saddle point caused by this symmetry, one can flip the sign of half of these matrix elements. Specifically, for eDSCn, for every $0<j<k\leq M^c$, we apply
\begin{align}
    \left(\tilde{\bm{F}}^{k\beta}\right)_{N+j,N+k} \longrightarrow -\left(\tilde{\bm{F}}^{k\beta}\right)_{N+j,N+k},
\end{align}
and for hDSCn, for every $0<k<j\leq M^c$, we apply
\begin{align}
    \left(\tilde{\bm{F}}^{j\beta}\right)_{N+1-j,N+1-k} \longrightarrow -\left(\tilde{\bm{F}}^{j\beta}\right)_{N+1-j,N+1-k}.
\end{align}
This procedure should be done before building $\bm{M}^j$ in Eqs. \ref{eq:m0edscn}-\ref{eq:mjedscn} and Eqs. \ref{eq:m0hdscn}-\ref{eq:mjhdscn}.

\subsection{Flowchart}\label{sec:flowchart}
For the convenience of the reader, let us now provide a flowchart of the final procedure. This flowchart is practically identical to the flowchart in Ref. \citenum{Qiu:2024:diis}
\begin{enumerate}[{\bf Step 1:}]
    \item Compute $\{\bm{Q}^{j>0}\}$ from Eq. \ref{eq:qj}, choose an initial guess $\bm{C}_{\rm ini}$ for $\bm{C}$ and initial guesses for $\{\lambda_j\}$, and choose a DIIS threshold $T_{D}$.
    \item Build density matrices $\{\bm{P}^j\}$ from $\bm{C}$ using Eq. \ref{eq:pj_from_C}.
    \item Build $\{w_j'\}$, $\{\bm{F}^{j\alpha}\}$, and $\{\bm{F}^{j\beta}\}$ from $\{\bm{P}^j\}$ using Eqs. \ref{eq:wp_2}-\ref{eq:wp_1} and \ref{eq:eje}, then build $\{\bm{M}^j\}$ using Eqs. \ref{eq:m0edscn}-\ref{eq:mjedscn}.
    \item Compute the initial DIIS error vector $\bm{V}^0$ using Eq. \ref{eq:diis_err_vec}, i.e.,
    \begin{align}
        \bm{V}^0 = [\bm{M}^0,\bm{P}^0] + \sum_{j=1}^{M^c} \left[\bm{M}^j-\sum_{k=1}^{M^f-1}\lambda_k\bm{Q}^k,\bm{P}^j\right].
    \end{align}
    \item If norm $(\bm{V}^0) < T_D$, jump to {\bf Step} \ref{step:diis_stop}. Otherwise, set the initial DIIS variable $\bm{A}^0$ to $\bm{0}$.
    \item Solve the nSCF problem for the new $\bm{C}$ and $\{\lambda_j\}$ using SQP with Eqs. \ref{eq:sqp_df}-\ref{eq:sqp_dg} given $\{\bm{M}^j\}$, and $\{\bm{Q}^{j>0}\}$. The nSCF threshold for the norm of the gradient and the constraint is set to norm $(\bm{V}^0)/100$.
    \item (DIIS iteration starts. Let the iteration counter $n$ starts from 1.) Build density matrices $\{\bm{P}^j\}$ from the new $\bm{C}$ using Eq. \ref{eq:pj_from_C}.\label{step:diis_start}
    \item Build $\{w_j'\}$, $\{\bm{F}^{j\alpha}\}$, and $\{\bm{F}^{j\beta}\}$ from $\{\bm{P}^j\}$ using Eqs. \ref{eq:wp_2}-\ref{eq:wp_1} and \ref{eq:eje}, then build $\{\bm{M}^j\}$ using Eqs. \ref{eq:m0edscn}-\ref{eq:mjedscn}.\label{step:build_fock}
    \item Calculate the DIIS error vector at iteration $n$:
    \begin{align}
        \bm{V}^n = [\bm{M}^0,\bm{P}^0] + \sum_{j=1}^{M^c} \left[\bm{M}^j-\sum_{k=1}^{M^f-1}\lambda_k\bm{Q}^k,\bm{P}^j\right].
    \end{align}
    \item Solve the nSCF problem for the new $\bm{C}$ and $\lambda$ using SQP with Eqs. \ref{eq:sqp_df}-\ref{eq:sqp_dg} given $\{\bm{M}^j\}$, and $\{\bm{Q}^{j>0}\}$. The nSCF threshold for the norm of the gradient and the constraint is set to norm $(\bm{V}^n)/100$.\label{step:nscf}
    \item Calculate the DIIS variable at iteration $n$:
    \begin{align}
        \bm{A}^n = \log\left(\bm{C}_{\rm ini}^\top\bm{C}\right).
    \end{align}
    \item If norm $(\bm{V}^n) < T_D$, exit the iteration and jump to {\bf Step} \ref{step:diis_stop}. Otherwise, use DIIS algorithm to build the interpolated DIIS variable $\bm{A}^{\rm next}$ from DIIS errors $\{\bm{V}^{i=0\cdots n}\}$ and previous DIIS variables $\{\bm{A}^{i=0\cdots n}\}$.
    \item Calculate the new $\bm{C}$ as
    \begin{align}
        \bm{C} = \bm{C}_{\rm ini}e^{\bm{A}^{\rm next}}.
    \end{align}
    \item Return to {\bf Step} \ref{step:diis_start}.
    \item DIIS convergence achieved.\label{step:diis_stop}
\end{enumerate}

\section{A More General Form of Constraints}\label{sec:general_constraint}
Before presenting results for the method above, we note that   the framework presented above 
around the constraints in Eqs. \ref{eq:constraints} and \ref{eq:constraints_v2}, 
can be applied to a more general set of constraints. In particular, for constraints written in the form of Eq. \ref{eq:trQP}, 
the full procedure discussed in Sec. \ref{sec:edsc} and \ref{sec:hdsc} remains unchanged if one generalizes
the kernel of the $j$th constraint to be
\begin{align}
    \bm{Q}^j = \sum_{i=1}^{M^f}q_{ij}\hat{P}^i,\label{eq:Q_general}
\end{align}
 Note that Eq. \ref{eq:Q_general} reduces to the constraints defined in Sec. \ref{sec:edsc} and \ref{sec:hdsc} if one fixes:
\begin{align}
    q_{ij}=\begin{cases}
        1, &i=j,\\
        -1, &i=j+1,\\
        0, &{\rm otherwise.}
    \end{cases}
\end{align}

\section{Numerical results}\label{sec:result}
\begin{figure*}[ht]
    \centering\includegraphics[width=0.65\textwidth]{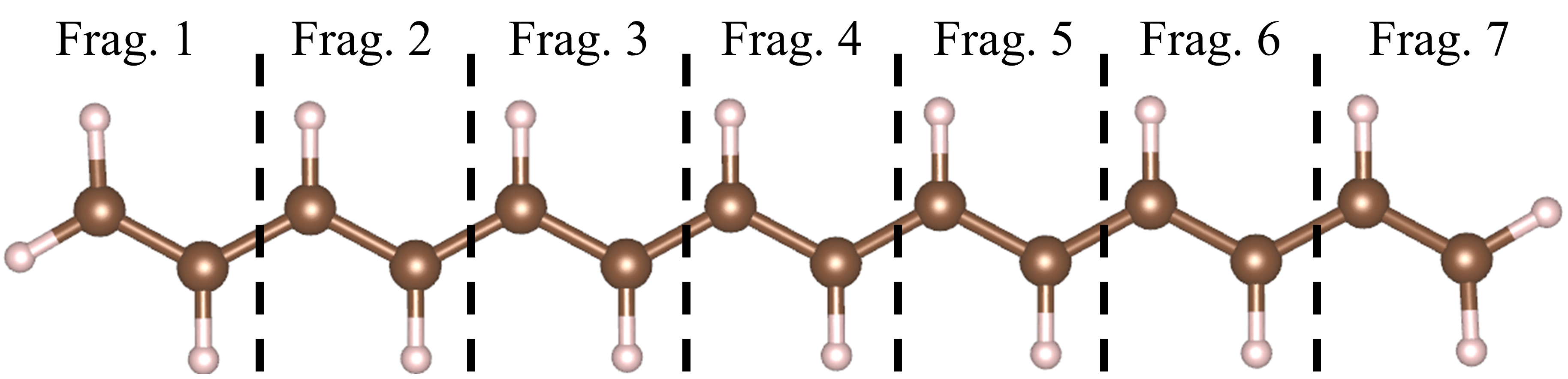}
    \caption{Geometry and SSH-chain with 14 carbon atoms and the partition of fragments.}
    \label{fig:ssh_c14}
\end{figure*}

We have applied the method above to a finite polyacetylene CH$_2$(CH)$_n$(CH$_2$)$^+$ cation chain, designed to mimic the famous SSH model problem. Here, we define each fragment to be a carbon dimer together with the relevant connected hydrogen atoms. A scheme of such a partition for the SSH chain with 14 carbon atoms (which will be referred to as ssh-c14) is shown in Fig. \ref{fig:ssh_c14}, where we have 7 fragments in total. For a SSH chain with $2M$ carbon atoms (ssh-2M), we partition the molecule into $M$ fragments in the similar way, and we include $M$ active orbitals within a  hDSCn (hole transfer) approach.

\begin{figure*}[ht]
    \centering\includegraphics[width=0.85\textwidth]{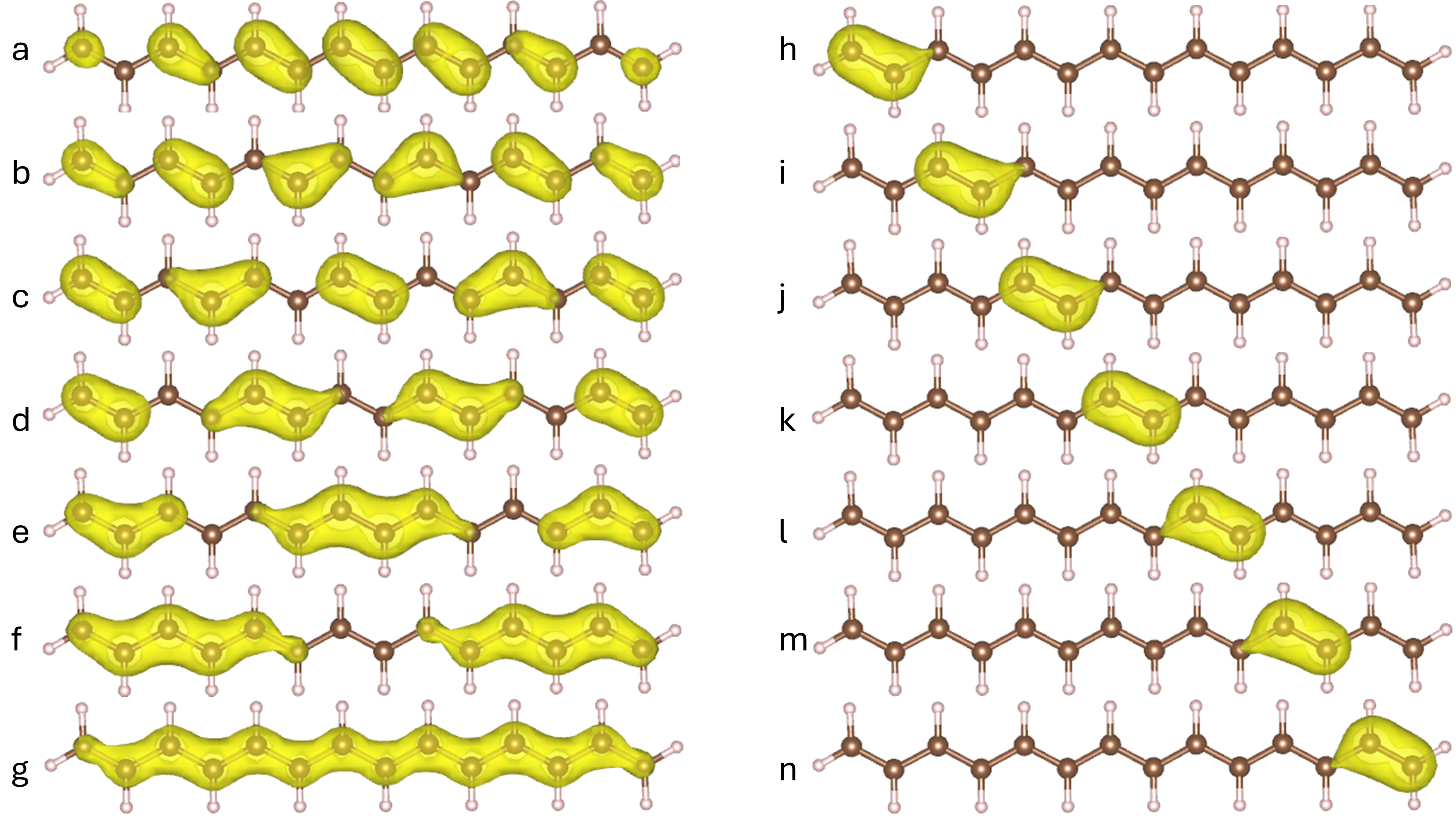}
    \caption{Charge density of active orbitals before (a-g) and after (h-n) Boys diabatization for SSH-chain with 14 carbon atoms.}
    \label{fig:adiabats_diabats}
\end{figure*}

Though perhaps not surprising to those who have worked with polyacetylene chains\cite{Paloheimo:1993:ssh,Hu:2015:ssh,Barborini:2015:ssh}, we have found that working with polyacetylene chains (as in Fig. \ref{fig:ssh_c14}) requires a bit of care. In particular, we have found that Eq. \ref{eq:constraints} is not in fact appropriate for this model: if we impose a constraint dictating that the vector space spanned by all active orbitals projects equally onto each molecular fragments, we find that hDSCn will ultimately converge to unwanted active orbitals due to boundary effect on the finite chain.   Nevertheless,  a meaningful, well-balanced set of  active orbitals can be defined by requiring that vector space spanned by all active orbitals projects equally onto the first fragment and the last fragment, then projects equally onto the second fragment and the second last fragment, and so forth.  For example, for ssh-c14, we impose three constraints of the form (using Eq. \ref{eq:Q_general}):
\begin{align}
    \bm{Q}^1 &= \hat{P}^{\rm frag. 1}-\hat{P}^{\rm frag. 7},\\
    \bm{Q}^2 &= \hat{P}^{\rm frag. 2}-\hat{P}^{\rm frag. 6},\\
    \bm{Q}^3 &= \hat{P}^{\rm frag. 3}-\hat{P}^{\rm frag. 5}.
\end{align}
Below, we report results for 
analogous SSH chains with $8, 10, 12,..., 26, 28$  carbon atoms.
Fig. \ref{fig:adiabats_diabats}a-g shows the density plot of the active orbitals.

\begin{figure*}[ht]
    \centering\includegraphics[width=0.9\textwidth]{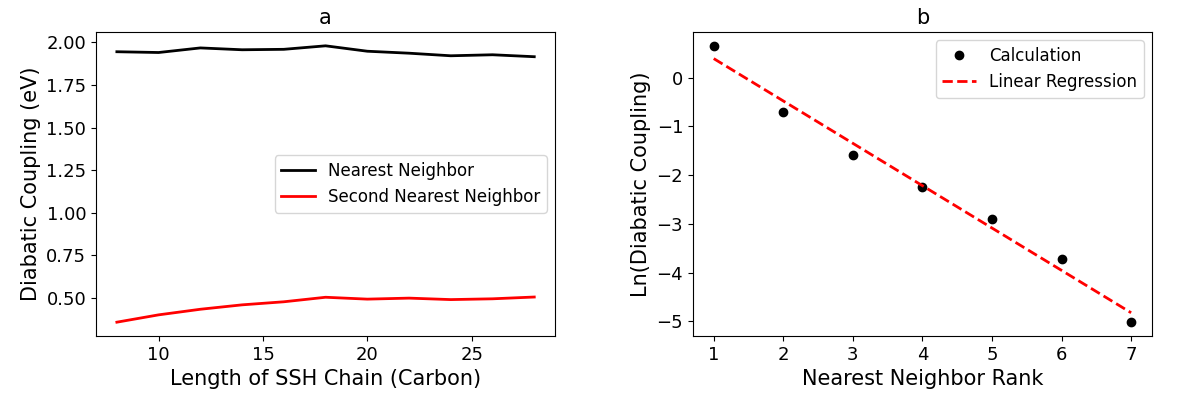}
    \caption{Diabatic coupling between orbitals. (a) Diabatic coupling between nearest and second nearest fragment as a function of the length of the SSH-chain. (b) Logarithm of diabatic coupling versus the rank of nearest neighbor in SSH-chain with 28 carbon atoms.}
    \label{fig:diabatic_coupling}
\end{figure*}

Next, in order to connect the present calculations to a true SSH model Hamiltonian, we 
applied Boys diabatization scheme \cite{subotnik:2008:boysgmh} to the active orbitals.  Figs. \ref{fig:adiabats_diabats}h-n portrays the diabatized orbitals.  The localized orbitals in Figs. \ref{fig:adiabats_diabats}h-n represent the carbon-carbon $\pi$-bond, as expected.
Finally, the most essential quantities to compute are the diabatic couplings between diabatic orbitals.  To that end, our results are presented  in Fig. \ref{fig:diabatic_coupling}. In Fig. \ref{fig:diabatic_coupling}a, we show the couplings between the nearest and the second nearest diabatic orbitals as a function of the length of the SSH chain; as one would hope, both quantities reach a constant plateau as the length of the SSH chain increases. In Fig. \ref{fig:diabatic_coupling}b, we further plot the logarithm of the diabatic coupling versus the rank of nearest neighbor for ssh-c28. The reasonably well linear relation shown here suggests the expected exponential decay of the diabatic coupling as a function of the distance. The $\beta$ value is 0.36 \AA$^{-1}$.

At the end of the day, it is clear that the present approach has delivered a meaningful means to parameterize an SSH model. Moreover, all of the data here arises from a smooth parameterization; one can easily begin to deform the nuclear geometry and investigate the resulting changes in diabatic couplings. To that end, in a companion paper, we will derive the relevant nuclear gradients and derivative couplings and begin exploring eDSC/hDSC potential energy surfaces.

\section{Discussion: A Trick for Convergence}\label{sec:discussion}

Before concluding this manuscript, a few words are appropriate regarding 
the flipping sign technique introduced in Sec. \ref{sec:saddle}. One cannot emphasize enough that the current algorithm quickly loses its efficiency without such a  sign flip and, as such, the physical interpretation of this procedure deserves further elaboration.

From a bird's eye point of view, the need for sign flip arises from the specific form of the configuration interaction (CI) Hamiltonian within eDSCn/hDSCn. Note that, for a standard CASSCF procedure, one usually optimizes the orbitals first and then diagonalizes the CI Hamiltonian to find the correct CI coefficients.
For our purposes, however, we have no such two-step procedure, but rather a one-shot optimization of the orbitals. 
Let us index the active orbital of the $j$th configuration as $a_j$. Then, the off-diagonal of the CI Hamiltonian gives
\begin{align}
    \bra{\Psi_j}\hat{H}\ket{\Psi_k} = \left(\tilde{\bm{F}}^{j\beta}\right)_{a_j,a_k}.
\end{align}
At the same time, however, 
 Eqs. \ref{eq:f_equal_e} and \ref{eq:f_equal_h} can be written together as
\begin{align}
    \left(\tilde{\bm{F}}^{j\beta}\right)_{a_j,a_k} = \left(\tilde{\bm{F}}^{k\beta}\right)_{a_j,a_k},\label{eq:f_equal}
\end{align}
Thus, the correct (converged) CASSCF solution therefore requires that several matrix elements of the Fock matrices be zero upon convergence. 

However, consider now the DIIS error vector in Eq. \ref{eq:diis_err_vec}; this vector  is essentially the gradient of the dynamically weighted sum of the energies (plus the Lagrange multiplier weighted sum of the constraints) w.r.t. the rotation generator. Upon convergence, the DIIS error vector should satisfy:
\begin{align}
    w_j'\left(\tilde{\bm{F}}^{j\beta}\right)_{a_j,a_k} -w_k'\left(\tilde{\bm{F}}^{k\beta}\right)_{a_j,a_k} = 0. \label{eq:diis_result}
\end{align}
Given Eq. \ref{eq:f_equal}, the above equation has two solutions, namely  either $w_j' = w_k'$ or $\left(\tilde{\bm{F}}^{j\beta}\right)_{a_j,a_k} = 0$. Note that the latter solution is the correct CASSCF solution while the former solution is the saddle point that we aim to avoid. By flipping the sign for one of the term in each pair, the saddle point solution is removed and we always reach the correct CASSCF solution.

\section{Conclusion}\label{sec:conclusion}

We have successfully extended the eDSC/hDSC methodology to treat charge transfer through multiple molecular fragments, developing a so-called eDSCn/hDSCn for systems with $n$ (i.e. more than two) charge centers. The method employs dynamically-weighted state averaging with constraints in a very general form. The DIIS-SQP algorithm developed previously\cite{Qiu:2024:diis} can be extended and remains extremely efficient for this constrained optimization problem. Moreover, the flexible constraint formulation allows adaptation to different molecular topologies and boundary conditions, an application to SSH chains demonstrates the method's ability to treat systems with strong couplings between charge localized diabatic states.
This extension opens new possibilities for studying complex charge transfer processes in extended systems. In particular, by prioritizing computational efficiency, the present approach is suitable for nonadiabatic dynamics simulations of CT across complex extended systems, and in a companion paper , we develop analytic gradients and derivative couplings for dynamics applications. 

Looking forward, beyond applications, given the current intense focus on chiral induced spin selectivity\cite{Bloom:2024:ciss} and spin-dependent electron transfer, future theoretical developments should focus on extensions of eDSC/hDSC to include complex orbitals with spin-orbital couplings (SOC) and a phase space Hamiltonian\cite{Tao:2024:vcd,Tao:2025:phasespace,Bradbury:2025:phasespace}; in particular, if one can successfully incorporate SOC and a phase space couplings, one hopes to learn a great deal about angular momentum conservation in the context of electron transfer. Finally, while the present approach is limited to one electron or one hole, one would like to explore closed-shell systems with double excitations. While  the computational cost would presumably increase dramatically when considering double excitations, some of our preliminary work has suggested that one can dramatically speed up constrained CASSCF(2,2) calculations  by folding CI diagonalization into the DIIS-SQP orbital optimization procedure for one-shot optimization cycle (as above). There are indeed still many open avenues for exploration as we seek the fastest means possible to coupled coupled nuclear-electronic charge transfer dynamics.  

\section{Acknowledgment}

\bibliography{cite.bib}
\end{document}